\documentstyle[prb,aps,psfig]{revtex}

\begin{document}

\title{Diluted Magnetic Semiconductors in the Low Carrier Density
Regime}
 
\author{R. N. Bhatt$^1$, Mona Berciu$^1$, Malcolm P. Kennett$^2$ and
Xin Wan$^3$}

\address{$^1$Department of Electrical Engineering, Princeton
University, Princeton, New Jersey 08544, USA.}

\address{$^2$Department of Physics, Princeton University, Princeton,
New Jersey 08544, USA.}

\address{$^3$Department of Physics, Florida State University,
Tallahassee, FL 32306, USA.}

\date{\today}

\twocolumn[\hsize\textwidth\columnwidth\hsize\csname@twocolumnfalse\endcsname

\maketitle
\begin{abstract}
This paper, based on a presentation at the Spintronics 2001
conference, provides a review of our studies on II-VI and III-V
Mn-doped Diluted Magnetic Semiconductors. We use simple models
appropriate for the low carrier density (insulating) regime, although
we believe that some of the unusual features of the magnetization
curves should qualitatively be present at larger dopings (metallic
regime) as well. Positional disorder of the magnetic impurities inside
the host semiconductor is shown to have observable consequences for
the shape of the magnetization curve.  Below the critical temperature
the magnetization is spatially inhomogeneous, leading to very unusual
temperature dependence of the average magnetization as well as
specific heat. Disorder is also found to enhance the ferromagnetic
transition temperature. Unusual spin and charge transport is implied.
\end{abstract}
\pacs{} ]

\section{Introduction}

Diluted magnetic semiconductors (DMS) are comprised of an inert host
semiconductor doped with both localized spins and 
 carriers (electrons or holes) that are either itinerant, or localized
on a much longer length scale.  In that sense, they belong to the
general family of correlated electron systems, which include a number
of fascinating materials such as cuprates, manganites, heavy fermions
and other Kondo lattice systems.

Electronic materials containing local moments have been studied for
some time. What makes the DMS so fascinating is that they belong to a
regime that has previously been neglected. While the name diluted
magnetic semiconductors implies (correctly) that the system has only a
small percentage of localized spins, they are at the opposite extreme
of the dilute magnetic alloys such as Fe or Mn in Cu, the canonical
systems involving itinerant fermion and localized spin degrees of
freedom, which have been studied extensively.\cite{OngBhatt} In the
dilute magnetic metallic alloys, the low density of spins are a
perturbation on the Fermi liquid representing the non-magnetic host
metal, so depending on the concentration of the local moments, they
may be studied in terms of dilute Kondo systems, or amorphous magnetic
systems with a spin-spin coupling mediated by the Fermi sea of
conduction electrons (RKKY coupling), which lead often to spin glass
behavior.\cite{Mydosh}

By contrast, in the regime of interest, the carrier density in DMS is
significantly {\em lower} than the (low) localized moment density, so
the spins become an integral part of the description of the system and
its magnetic phase, rather than a mere perturbation on a metallic
Fermi sea. In that sense, the situation is even more extreme than {\em
e.g.}, in Kondo lattice and heavy Fermion materials, where the two
species have comparable densities. This large, inverted, ratio of
local moments to carriers is in fact similar to that in the high $T_c$
cuprates. However, unlike the cuprates, the density of local moments
is low and incommensurate with the lattice, and the carriers and the
spins are not in the same band.  As a consequence of the low moment
density, the exchange between local moments is not standard direct or
superexchange, as in the cuprates, but is mediated by the carriers,
even though their density is so small.  Thus, the DMS are in rather
different region of phase space of electronic materials with local
moments, than other correlated electron systems.

Despite this difference, most models of diluted magnetic
semiconductors start from the high carrier density limit, where the
carriers may be modeled as free carriers moving in the conduction or
valence band.\cite{MacDonald-Dietl} This is understandable, since in
the high density limit the carrier kinetic energy is the largest
energy in the problem, and calculations may be done perturbatively
starting from the non-interacting Fermi gas. However, most of the
interesting behavior is seen at low carrier densities, where the
system is insulating, or not too far from the metal-insulator
transition.  Consequently, we have concentrated in this work on the
low density regime, starting from bound carriers, and moving on to
carriers in an impurity band formed from the bound impurity states.

As we wish to cover the case of insulating behavior at arbitrary
filling factor {\em i.e.,} away from the half filled impurity band
case (one carrier per site), disorder has to be included at the
outset. In particular, we model the system with randomly distributed
dopants, as in the experimental system, since it has been recognized
that the random distribution is essential to understand the magnetic
properties of conventional, non-magnetic doped
semiconductors.\cite{BhattLee,Bhatt} Such models exhibit both
insulating and metallic phases, and with the random distribution of
impurity sites included, are in a position to reveal the effect of
disorder in low carrier density systems. In the case of the
predominantly antiferromagnetic couplings between hydrogenic centers
in conventional doped semiconductors, the randomness is found to
suppress magnetic order below measurable temperatures ( $\sim$
millikelvin ), and possibly to zero. In the case of DMS, where
interactions lead to ferromagnetic ordering,\cite{pwolff,Angelescu} in
agreement with experimental findings,\cite{bookIIVI,Ohnorev} we find
that randomness leads to unusual behavior in the magnetic response,
and effects of randomness are expected in the transport behavior as
well.

In this paper, we review the results of our approach to DMS based
on both II-VI semiconductors (like CdTe or ZnSe), and on III-V
semiconductors (such as GaAs or GaP), and compare the two families of
DMS systems.  The paper is organized as follows. In Section \ref{sec2}
we briefly review the properties of conventional (insulating)
ferromagnets. The results presented serve as a reference with which to
contrast the results obtained in the rest of the paper. Section
\ref{sec3} addresses the II-VI based DMS, while Section \ref{sec4}
deals with III-V DMS. In each case, we introduce the Hamiltonians we
use to model these systems. Results obtained within mean-field
approximation and with Monte Carlo simulations are presented.  The
effect of positional disorder of the Mn dopants is studied, as are the
similarities and differences between II-VI and III-V DMS. Finally,
Section \ref{sec5} summarizes our results and conclusions. It also
includes a discussion of important issues such as robustness of
models, relevance of disorder on spin scattering, and key experiments
which could help provide a better understanding of these materials.

\section{Conventional Ferromagnetic Systems}
\label{sec2}

A typical model of a uniform ferromagnet consists of a collection of
identical magnetic spins of magnitude $S$, placed on an ordered
Bravais lattice. While the generic case may be anisotropic in spin
space due to a variety of reasons (spin-orbit coupling, crystal fields
etc.) we consider here the simplest isotropic case where the spin
interactions are well described by a Heisenberg Hamiltonian
\begin{equation}
\label{1}
{\cal H}=-\sum_{i\ne j}^{} J_{ij} \vec{S}_i \cdot \vec{S}_j - \vec{H}
\cdot \sum_{i}^{} g \mu_B \vec{S}_i.
\end{equation}
Due to translational invariance, the exchange integral $J_{ij}$
depends only on the distance $\vec{R}_i-\vec{R}_j$ between spins. In
insulating materials, this dependence is due to overlap between the
electronic orbitals involved in creating the spin S (through Hund's
rule), leading to an exponential decay of $J_{ij}$ with increasing
distance.  In typical magnetic atoms, these orbitals are of $d$ or $f$
type, and they are localized within $\sim 1-2~$\AA of the nucleus.  As
a result, it is customary to restrict the first sum in Eq. (\ref{1})
to only nearest-neighbor spins. The external magnetic field $\vec{H}=H
\hat{e}_z $ breaks the rotational symmetry, leading to the appearance
of a non-vanishing expectation value $\langle S^z_i \rangle $ at each
site. Translational invariance implies that $\langle S^z_i \rangle =
\langle S\rangle$ is independent of the position $\vec{R}_i$ of the
spin.

\begin{figure}
\centering
\parbox{0.49\textwidth}{\psfig{figure=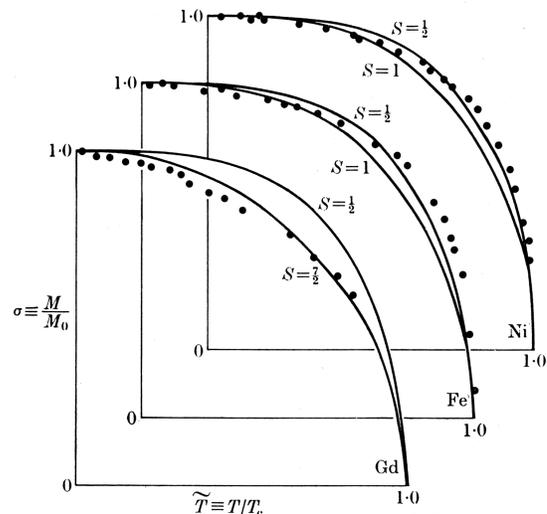,width=80mm}}
\caption{\label{fig2} Dependence of reduced magnetization $M/M_0$ upon
reduced temperature $T/T_C$. Curves are slightly different for
different values of the quantum spin $S$, however they all have a
convex upward shape.  The solid circles represent typical experimental
data for Gd ($S\approx {7 \over 2}$), Fe ($S\approx 1$) and Ni
($S\approx {1 \over 2}$). (From Stanley\cite{stanley}). }
\end{figure}

\begin{figure}
\centering
\parbox{0.49\textwidth}{\psfig{figure=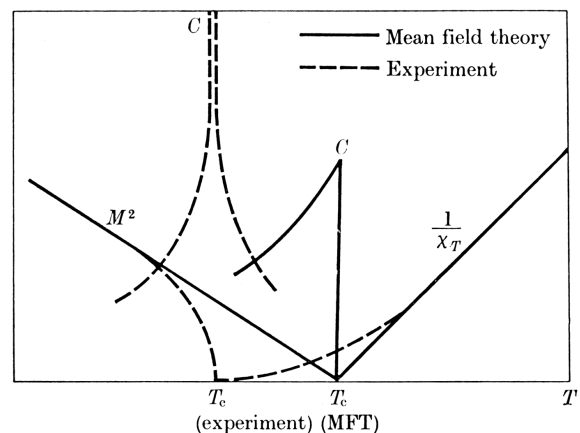,width=80mm}}
\caption{\label{fig3} Schematic comparison of typical experimental
measurements of temperature dependence of magnetization, specific heat
and susceptibility for a Heisenberg ferromagnet (such as EuS) with the
predictions of the Weiss mean-field theory. Note that the curve for
inverse susceptibility $1/\chi_T$ is shown only for $T>T_C$. (From
Stanley\cite{stanley}). }
\end{figure}

While an exact solution for the Heisenberg Hamiltonian (\ref{1}) is
known only in one dimension, it has been found that the Weiss
(mean-field) approximation provides a qualitatively good understanding
of the properties of these systems. The mean-field factorization
$\vec{S}_i \cdot \vec{S}_j \rightarrow \vec{S}_i \cdot
\langle\vec{S}_j\rangle + \langle \vec{S}_i \rangle \cdot \vec{S}_j -
\langle\vec{S}_i \rangle \cdot \langle \vec{S}_j \rangle = \langle
S\rangle \left( S_i^z + S_j^z\right) - \langle S\rangle^2$ allows for
a solution of the problem in terms of an effective magnetic field
$H(i) = H + J\langle S\rangle/(g\mu_B) $, where $J=\sum_{j \ne i}
J_{ij}$. [If only nearest-neighbor interactions are kept, $J=z
J_{01}$, where $z$ is the coordination number of the Bravais lattice
and $J_{01}$ is the exchange integral for nearest-neighbor spins].  In
the absence of an external magnetic field, a non-vanishing solution
for $\langle S\rangle$ is found for $T \le T_C = J S(S+1)/3k_{\rm B}
$. In other words, the system is ferromagnetically aligned below the
critical temperature $T_C$, and the spontaneous magnetization $\langle
M \rangle = g \mu_B \langle S^z_i\rangle$ increases rapidly
(Fig. \ref{fig2}) with decreasing temperature and is already close to
the saturation value $M_0 = g \mu_B S$ below $T < 0.5 T_C$.
Concurrently, the specific heat has a peaked structure around $T_C$
and drops rapidly to zero for $T < 0.5 T_C$ reflecting the fact that
the only accessible degrees of freedom for low T are the
long-wavelength (collective) spin-wave excitations which have a
restricted phase space (see Fig. \ref{fig3}).\cite{stanley}

It is well-known that mean-field approximations overestimate the
strength of the correlations, leading to rather high estimates for the
Curie temperatures $T_C$. Detailed studies of these Hamiltonians with
Monte-Carlo simulations, which properly account for the effects of
thermal fluctuations, find quantitative changes of up to a factor of 2
in the value of $T_C$. However, as suggested in Fig. \ref{fig3}, the
{\em qualitative} features of the magnetization, specific heat and
susceptibility curves remain as in the Weiss mean-field treatment, in
good agreement with experimental measurements.

\section{Ferromagnetic systems: II-VI DMS}
\label{sec3}

\subsection{The model}

The II-VI DMS are based on semiconductors AB, where A is a group-II
element and B is a group-VI element (such as CdTe or ZnSe). In the
II-VI DMS, some of the divalent sites (Cd/Zn) are substituted by a
magnetic element, typically Mn.
{\em This fraction is denoted by $x$}, so the DMS we consider is
A$_{1-x}$Mn$_x$B.  Mn is also a group-II element, but in addition it
has a half-filled $3d$ shell, with a total spin given by Hund's rule:
$S = 5/2$. In the absence of other types of dopants, the system
A$_{1-x}$Mn$_x$B is an insulator which exhibits antiferromagnetic
tendencies at low temperatures. This is seen, for instance, from
measurements of the susceptibility which is found to depend on
temperature as $\chi(T) \sim 1/(T+T_N)$, with a Neel temperature of a
few kelvin. \cite{bookIIVI,Mike} The origin of this antiferromagnetic
(AFM) tendency is the (expected) antiferromagnetic exchange between
the Mn spins. However, for low doping concentrations $x$, the average
distance between Mn spins is large and this AFM direct exchange is
rather small.

When a low density of charged dopants, such as group-V Phosphorus (P)
substituting for the group-VI element, is introduced in the system,
each of them binds a hole (or electron) in a shallow hydrogenic $1s$
state $\phi(\vec{r})\sim \exp{\left(-r/a_B\right)}$, characterized by a
Bohr radius $a_B\sim 10-20$ \AA. Exchange interactions arise between
the spins of these charge carriers and the Mn spins, and are described
by the Hamiltonian\cite{pwolff}:
\begin{equation}
\label{3.1}
H = \sum_{i,j} J({\vec r}_i, {\vec R}_j) {\vec s}_i \cdot {\vec S}_j.
\end{equation}
Here, $\vec{S}_j$ is the spin of the Mn at position ${\vec R}_j$ and
${\vec s}_i$ is the spin of the electron/hole centered at ${\vec
r}_i$. The exchange interaction $J({\vec r}_i, {\vec R}_j)$ is
dependent on the overlap between the orbital $\phi(\vec{r}-\vec{r}_i)$
of the charge carrier and the orbitals $\psi_d(\vec{r}- \vec{R}_j)$ of
the $3d$ electrons responsible for the Mn spin. Since these $3d$
orbitals are localized on a scale of a few \AA $~$ around the Mn nucleus,
the exchange is proportional to the carrier charge density at the Mn
site, {\em i.e.}
\begin{equation}
\label{3.2}
J({\bf r}_i, {\bf R}_j) = J_0 |\phi(\vec{R}_j-\vec{r}_i)|^2=J_0e^{- 2
|{\bf r}_i - {\bf R}_j| / a_B},
\end{equation}
where $J_0$ characterizes the strength of the exchange. Typically, for
electrons $J_0 <0$, while for holes $J_0>0$. However, since in the
following we treat the spins as classical variables, the sign is
irrelevant. For specificity, in the rest of the paper we assume
$J_0>0$ corresponding to holes as charge carriers.

The Hamiltonian (\ref{3.1}) neglects the direct AFM interactions
between the Mn spins. For low values of the fraction $x$, it can be
simply accounted for in the following manner. For Mn spins which are
very close to one another (such as nearest neighbors), the direct AFM
exchange is the dominant (large) interaction, and leads to the
formation of a singlet state. This singlet becomes inert as far as
magnetic interactions are concerned. For Mn spins which are fairly far
apart from other Mn spins, the dominant magnetic interaction is the
exchange with the charge carrier spins. As a result, to first order
the Hamiltonian (\ref{3.1}) accounts for both types of interactions if
we restrict the summation over the Mn spins to only those Mn spins
which are not part of a spin-singlet. At low $x$, this
includes a large majority of Mn spins.  If the fraction $x$ of Mn
becomes too large, both types of interactions will be of comparable
size for all the Mn spins, and therefore this separation is no longer
possible. In this case, the frustration imposed by the competing
exchanges leads to the appearance of a spin-glass state, which has
been observed experimentally for $x \ge 0.2$.\cite{spinglass} In the
following, we restrict ourselves to the low $x$ ($ x \le 0.1 $) limit.

\begin{figure}
\centering
\parbox{0.45\textwidth}{\psfig{figure=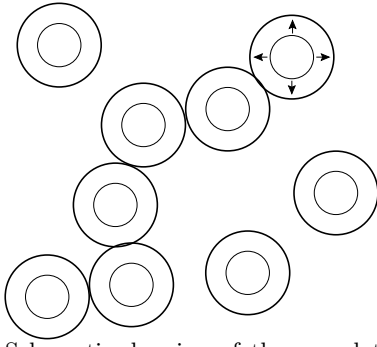,width=50mm,angle=270}}
\caption{\label{figdis} Schematic drawing of the percolation limit for
a disordered collection of BMPs. As the temperature is lowered, the
size of each BMP increases and a percolated network appears below
$T_C$. Just below $T_C$, only a small fraction of the spins belong to
the percolated network and sustain the bulk magnetization of the
sample. The large majority of spins is outside the percolated network
and behave like quasi-free (non-interacting) spins.  }
\end{figure}

Simple thermodynamic considerations show that, qualitatively, at a
temperature $k_BT < J(r)$ [see Eq. (\ref{3.2})], all Mn spins within
distance $r_T\sim \left(a_B/2\right) \ln{\left(J_0/k_BT\right)}$ of a
dopant order their spins antiferromagnetically with respect to the
dopant hole spin. As a result, a region with a large magnetization
(from all the parallel polarized Mn spins) appears near the
dopant. This is known as a Bound Magnetic Polaron (BMP),
\cite{bookIIVI} whose radius $r_T$ (see above) increases
logarithmically with decreasing temperature. As a result, one expects
that long-range ferromagnetic order appears in the system for
temperatures low enough that a continuous percolating network of BMPs
is formed (as shown schematically in Fig. \ref{figdis}), provided that
nearby BMPs prefer to orient ferromagnetically with respect to one
another. At first sight, this seems to not be the case, since direct
exchange between the charge carriers localized in hydrogenic orbitals
has an antiferromagnetic sign.\cite{BhattLee,Bhatt,Andres} However,
this antiferromagnetic coupling is overwhelmed by effectively
ferromagnetic interactions between BMPs coming in part from Mn spins
in between the polarons which favor ferromagnetic alignment of Mn
spins\cite{pwolff} and partly from the modification of the effective
direct exchange as a result of the local field due to the polarized
Mn.\cite{Angelescu}

These mechanisms favoring parallel orientation of the BMPs at low
temperatures are rather weak, and as a result the Curie temperature
below which long-range ferromagnetism is observed in these systems is
very low, to our knowledge below 5 K for all II-VI DMS studied so
far. Moreover, as the transition is of a percolation type, and the
percolation fraction is $\sim$~20\% for three dimensions, this implies
that just below $T_C$ about $~$80\% of the Mn spins do not participate
in the ferromagnetism. These are the Mn spins which are outside the
percolated cluster, i.e. far from the charged dopants (see
Fig. \ref{figdis}). They are very weakly interacting (essentially
disordered) unless the temperature becomes so low that a nearby BMPs
grows large enough to include them.  This results in a very unusual FM
phase, in which a substantial part of the spin entropy survives down
to very low T.

\subsection{ Monte Carlo simulations}

 We performed Monte Carlo simulations on the Hamiltonian (\ref{3.1}),
to study this unusual FM phase, treating both Mn and carrier spins as
classical variables.\cite{wan} This appears to be a reasonable
approximation, since $S=5/2$ is a large spin and the Mn spins dominate
the magnetic response. Simulations were carried out for zinc-blende
lattices with lattice constant $a=5$ \AA, for Mn concentration $x =
0.001$, dopant density $n_d = 10^{18}$~cm$^{-3}$ and $a_B = 20$
\AA. The exchange $J_0$ defines the unit of energy.  With these
parameters, the Mn concentration $n_{Mn}=4x/a^3$ is 32 times the
dopant concentration. Nevertheless, the magnetic coupling is mediated
by the latter because of the large Bohr radius, as required for the
polaron picture to hold.

\begin{figure}
\parbox{0.49\textwidth}{\psfig{figure=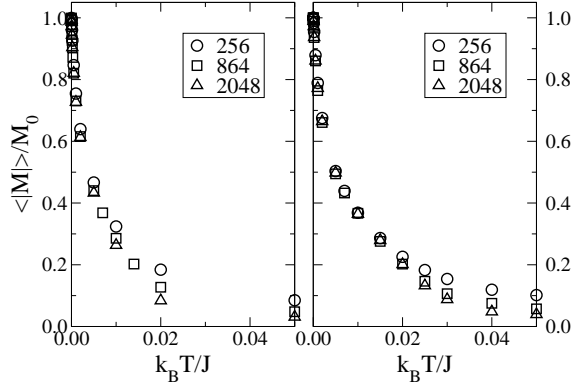,width=80mm,angle=270}}
\caption{\label{fig4} Magnetization per Mn spin as a function of
temperature, in a II-VI DMS, for classical/discrete spin model
(left/right panel). Results are shown for samples with $N=256,864$ and
2048 Mn spins. Finite size scaling analysis finds a critical
temperature $T_C=0.014 J_0$.\cite{wan} The magnetization curves are
very unlike the conventional ferromagnet magnetization curve shown in
Fig. \ref{fig2}. }
\end{figure}

The magnetization curves obtained have unusual, concave upward shapes
(see Fig. \ref{fig4}, left panel), very unlike the typical
magnetization curve of Fig. \ref{fig2}. For these parameters, the
critical temperature $T_C=0.014 J_0$ is found using finite size
scaling.  \cite{wan} We find that the magnetization reaches its
saturation value only at exponentially small temperatures, reflecting
the existence of the quasi-free Mn spins outside the percolated
(magnetically ordered) region.

\begin{figure}
\centering
\parbox{0.49\textwidth}{\psfig{figure=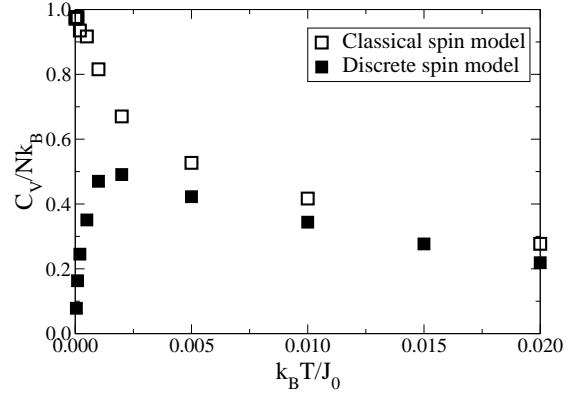,width=80mm,angle=270}}
\caption{\label{fig5} Specific heat per Mn spin as a function of
temperature, for classical (empty squares) and discrete (full squares)
spin models.  Systems with $N=2048$ Mn spins were used in both
cases. While the discrete model recaptures the proper limit $C_V
\rightarrow 0$ as $T \rightarrow 0$, the peak in the specific heat is
well below the critical temperature ($T_C=0.014J_0$ for these
parameters), unlike in conventional ferromagnets, where the peak in
the specific heat is at $T_C$ (see Fig. \ref{fig3}). }
\end{figure}

The specific heat of the classical Heisenberg model has the unphysical
limit $C_{\rm V} \rightarrow N k_{\rm B}$ as $T\rightarrow 0$ (empty
squares in Fig. \ref{fig5}). While this agrees with the equipartition
theorem, it implies that quantum mechanics (with discrete energy
levels) is needed to capture the correct limit $C_{\rm V} \rightarrow
0$ as $ T \rightarrow 0$.  One way to mimic the discretization, but
avoid the complexities of the quantum Monte Carlo treatment, is to use
a discrete (classical) vector model, in which each Mn spin can only be
oriented along one of the six [100] directions. An efficient Monte
Carlo method for this discrete model is described in
Ref. \onlinecite{wan}.  While the magnetization curves are very
similar to the ones obtained in the continuous spin Heisenberg model
(see Fig. \ref{fig4}), the specific heat results are very different
(see Fig. \ref{fig5}). As expected, for the discrete model $C_{\rm
V}\rightarrow 0$ as $T \rightarrow 0$. However, unlike in the case of
a typical FM, the peak in $C_{\rm V}$ is not near $T_C$, but at
temperatures well below $T_C$.  This reflects the residual entropy of
the free Mn spins outside the percolated region.

\subsection{Effect of disorder}

In II-VI DMS there are two sources of positional disorder: disorder in
the positions of the Mn spins and disorder in the position of the
charged dopants. In the limit when there are many Mn ions per dopant,
the Mn spin disorder is not expected to have a significant effect on
the magnetization curves, or the critical temperature. The reason is
that at the very low temperatures where percolation appears, the
radius of each BMP is significantly larger than $a_B$, favoring
interactions with a large number of Mn spins.  Disorder in the Mn
positions will lead to some fluctuations in the average number of Mn
spins found in each BMP, but this should have a relatively small
effect.

On the other hand, disorder in the position of the charged carriers
(centers of the BMPs) has a large effect on the critical
temperatures. As seen from Fig. \ref{figdis}, disorder in the
positions of the BMPs facilitates the appearance of a large percolated
cluster for smaller BMPs sizes (larger temperatures), since only a
subset of the BMPs must percolate in order for ferromagnetic order to
appear in the system.  On the other hand, the ordered BMP lattice only
percolates when each and every BMP is included. This obviously happens
when a larger fraction of the space is filled by BMPs, i.e. at a lower
temperature. However, we emphasize again that even for the ordered BMP
lattice, a significant volume containing a large fraction of the Mn
spins is still outside the percolated volume (in the interstitial
spaces) and therefore the phenomenology related to the existence of
weakly-interacting spins down to exponentially low temperatures is
still valid.

We have verified, using Monte Carlo simulations, that the critical
temperature of a system in which the charge carriers are placed in an
ordered superlattice is lower than that of a sample with random
positions of the charge carriers, when all other parameters are
identical. For the case investigated, the relative increase of $T_C$
with disorder was $50\%$.\cite{wan2} However, this relative increase
is expected to depend on the various parameters of the problem.

\section{Ferromagnetic systems coupled to fermions: III-V DMS}
\label{sec4}

\subsection{Introduction}

When Mn is doped in a III-V semiconductor, such as GaAs, the major
difference with respect to the II-VI DMS is that the Mn atom provides
both the $S=5/2$ spin and the dopant charge carrier (a hole, since
divalent Mn substitutes for trivalent Ga). While this implies
nominally equal concentrations of holes and Mn spins, experimentally
it is found that the hole concentration is only $p = 10-30\%$ of the
Mn concentration.\cite{Ohnorev,Esch} The  compensation process(es)
responsible for the removal of such a large fraction ($\sim 70-90\%$)
of the holes from the carrier band are not fully understood, but it is
believed that an important role is played by As antisite defects.
Such defects are created when group-V As substitutes for group-III Ga,
and removes two holes introduced by Mn impurities, thus effectively
decreasing the hole concentration. Compensation is responsible not
only for the substantial decrease of the hole concentration, but also
leads to the appearance of charged compensation centers
(e.g. As$^{2+}$ for As antisites). The Coulomb potential created by
these charged compensation centers may also play a role in the physics
of these systems, as we discuss in the following.

As in the (II,Mn)VI systems, the main magnetic interaction in the
(III,Mn)V DMS is the exchange between the Mn spins and the hole spins,
which is known to be antiferromagnetic.\cite{Ohnorev} Assuming, again,
very sharply peaked Mn $3d$ orbitals, this exchange is proportional to
the probability of finding the charge carrier at the Mn site. This
probability is extracted from the wave-functions of the orbitals
occupied by the hole charge carriers. The appropriate framework to
describe the hole states depends on their concentration.  At low hole
concentrations, screening processes are ineffective. The unscreened
Coulomb potentials of the Mn dopants are responsible for the splitting
of hydrogen-like impurity levels from the top of the valence band, and
the holes occupy these impurity states. In the limit of high hole
concentrations when the carrier kinetic energy is the largest energy
in the problem, the Coulomb potential of the Mn dopants effectively
gets reduced because of screening. As a result, the holes occupy a
Fermi sea at the top of the valence band. Qualitatively, it is
apparent that the two situations could lead to quite different
physics. Holes occupying Bloch states in the valence band are found
with equal probability anywhere inside the host semiconductor, and
therefore one expects the system to be rather homogeneous. On the
other hand, holes occupying impurity states are found with high
probability near the Mn sites. As a result, we expect a rather
inhomogeneous distribution of the holes in the host semiconductor, and
the positional disorder of the Mn dopants may play an important role,
since it defines the length-scale for these inhomogeneities.

Ga$_{1-x}$Mn$_x$As has a Metal-Insulator Transition for $x \sim 0.03$
and shows re-entrant insulating behavior for $x > 0.07$.\cite{Ohnorev}
In the insulating regimes, the low-temperature conductivity is
consistent with Mott long-range variable hopping, \cite{Esch,Katsu}
suggesting the existence of impurity-like levels. Even for the most
metallic sample ($x=0.053$) the screening length ($l \sim 10$ \AA) as
evaluated from the Thomas-Fermi theory is of comparable size, not much
smaller than the Bohr radius of the impurity level ($a_B \sim 8$
\AA).\cite{Bhatta,mona}

\subsection{The model}

Motivated by these observations, we have attempted to understand the
low $x$ regime within a model based on the existence of impurity
hydrogen-like orbitals at each Mn site. While this is similar to our
approach to the II-VI DMS systems, one difference is that since the
number of holes is smaller than the number of Mn, there must be a
mechanism to allow the holes to ``choose'' the Mn dopants near which
to stay. Such a mechanism is naturally provided by hopping processes
facilitated by the overlap between impurity wave-functions centered at
different Mn sites. Therefore, the Hamiltonian describing such a
system is of the form
$${\cal H}=\sum_{i,j}^{}t_{ij}c^{\dagger}_{i\sigma} c_{j\sigma}
+\sum_{i}^{}\left[u(i) c^{\dagger}_{i\sigma} c_{i\sigma} +
Un_{i\uparrow} n_{i\downarrow}\right]
$$ 
$$ +\sum_{i,j}^{} J_{ij} \vec{S}(i) \cdot {\vec s}_j + \sum_{i,j}^{}
{K}_{ij} \vec{S}(i)\cdot \vec{S}(j)
$$
\begin{equation}
\label{4.1}
- g \mu_BH\sum_{i}^{}{\sigma \over 2} c^{\dagger}_{i\sigma}
  c_{i\sigma} - \tilde{g} \mu_B H \sum_{i}^{}S^z(i).
\end{equation}
Here, $i$ indexes different Mn positions ${\bf R}_i$, and
$c_{i\sigma}^{\dagger}$ is the creation operator for a hole with spin
$\sigma$ in the impurity level centered at ${\bf R}_i$, while
$\vec{S}_i$ is the spin of the corresponding Mn dopant.

The first line in Eq. (\ref{4.1}) is the Hamiltonian of the charge
carriers. The first term describes hopping of holes between impurity
levels. For simplicity, we assume again 1$s$ impurity states with
$\phi(\vec{r})= \exp{\left(-r/a_B\right)}$. In fact, the hole impurity
wave-function is more complicated, due to the band-structure of the
valence band from which it splits (for details, see
Ref. \onlinecite{Bhatta}). For the hopping integral we use the simple
parameterization $t_{ij} = 2(1+r/a_{\rm B})\exp{(-r/a_{\rm B})}$ Ry,
where $r=|{\bf R}_i - {\bf R}_j|$, appropriate for hopping between two
isolated 1$s$ impurities which are not too close to one
another. \cite{Bhatt1} For Mn doped into GaAs, the Bohr radius is
$a_B=7.8$~\AA~ and the binding energy which defines the Rydberg is 1
Ry =110 meV.  \cite{Bhatta,mona} We have investigated other
parameterizations for the hopping matrix $t(r)$ elsewhere,\cite{MB}
and found that while they lead to quantitative changes, qualitatively
the results are similar.

The second term describes an on-site potential $u(i)$ due to the
Coulomb potential of the other Mn impurities, as well as other nearby
charged compensation centers. An on-site Coulomb repulsion $U$ of the
Hubbard type may be added to describe the electron-electron repulsion
between electrons occupying the same impurity orbitals. For isolated
1s impurities, $U \approx 1~$ Ry. However, depending on the
effectiveness of screening, the electron-electron interactions may be
longer-range. A fully self-consistent treatment of this problem should
involve a proper description of the screening processes, and would
allow a detailed computation of the strength of the hopping matrix,
the on-site Coulomb potential and the electron-electron
interactions. However, since the full self-consistent description is
extremely difficult to achieve, especially as details about
compensation processes are still not clarified, we use the simplified
assumptions described above. We believe that they should provide a
good qualitative description of the properties of these compounds, and
with proper fitting of various energy and length scales may even lead
to a quantitative description.

The second line of Hamiltonian (\ref{4.1}) describes the AFM exchange
between the Mn spin ${\vec S}_j$ and the hole spin ${\vec s}_i={
1\over 2} c^{\dagger}_{i\alpha} {\vec \sigma}_{\alpha\beta}
c_{i\beta}$ [${\vec \sigma}$ are the Pauli spin matrices]. As in II-VI
DMS, the AFM exchange is proportional to the probability of finding
the hole trapped at ${\vec R}_i$ near the Mn spin at ${\vec R}_j$, so
$J_{ij} = J \exp{\left(-2 |{\vec R}_i-{\vec R}_j|/a_B \right)}$. Based
on calculations\cite{Bhatta} of the isolated Mn impurity in GaAs, we
estimate the exchange coupling between a hole and the trapping Mn
($\vec{R}_i=\vec{R}_j$) to be $J=15$ meV.

The second term describes the direct Mn-Mn exchange in the
 semiconductor host, which is expected to be short range, and
 consequently not important at low $x$ when Mn are a few sites away
 from each other. We have therefore omitted this term ({\em i.e.} set
 $K_{ij}=0$); however, for higher concentrations this may be
 important.  Finally, the third line in Hamiltonian (\ref{4.1})
 describes the interaction with an external magnetic field.

Given the large number of terms in the Hamiltonian, it is useful to
try to understand the effect of each. To begin with, we neglect the
random on-site potential ($u(i)=0$), the electron-electron interaction
($U=0$), the direct Mn-Mn AFM interactions (${K}_{ij}=0$) and turn off
the external magnetic field ($H=0$) (we will discuss the effects of
these various terms later on).  As a result, the Hamiltonian contains
only its two main terms ($t_{ij}$ and $J_{ij}$), describing the
dynamics of the charge carriers and the AFM interaction between the Mn
spins and the charge carrier spins.

\subsection{Similarities and differences between II-VI and III-V DMS} 

\begin{figure}
\centering
\parbox{0.45\textwidth}{\psfig{figure=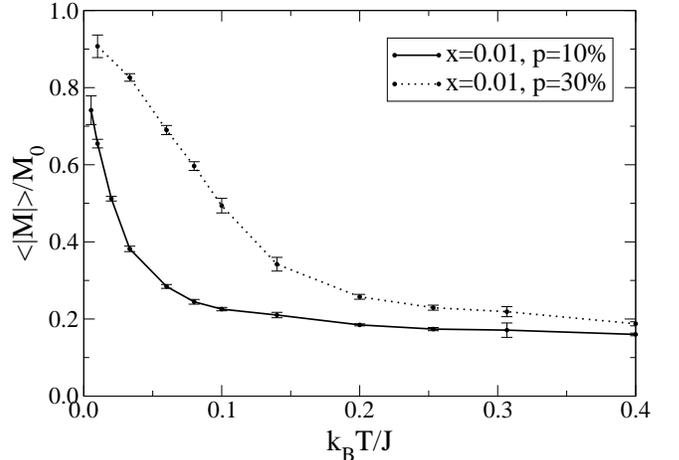,width=85mm,angle=270}}
\caption{\label{figMC1} Magnetization per Mn spin as a function of
temperature, in a III-V DMS, using Monte Carlo
simulations.\cite{malcolm} Curves correspond to $x=0.01$, and relative
hole to Mn concentrations $p=10$ and $ 30\%$. }
\end{figure}

\begin{figure}
\centering
\parbox{0.45\textwidth}{\psfig{figure=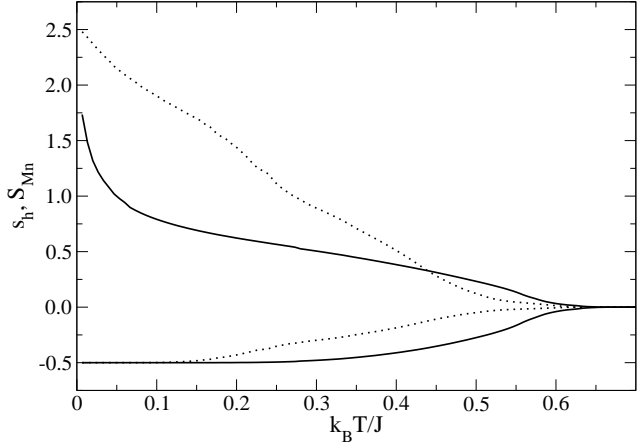,width=85mm,angle=270}}
\caption{\label{figN1} Average magnetization of the Mn spin ($S_{Mn}
>0$) and of the charge carrier spins $s_h <0$) as a function of
temperature, in a III-V DMS. Curves correspond to $x=0.0093$ and
$p=10$ (full line) and $ 30\%$ (dotted line), and were obtained using
the mean-field approximation.\cite{MB} }
\end{figure}

We investigated the Hamiltonian (\ref{4.1}) using both the mean-field
approximation (MFA)\cite{mona,MB} and Monte Carlo (MC)
simulations.\cite{malcolm} Typical magnetization curves obtained with
MC methods for a Mn concentration $x=0.01$ and hole concentrations
$p=10$ and $30\%$ are shown in Fig. \ref{figMC1}.  The corresponding
curves obtained using the mean-field approximation for the similar
parameters are shown in Fig. \ref{figN1}.  While there are
substantial quantitative differences between the two, these are easily
understandable. The long tail of the MC curves at high $T$ are due to
finite sizes of the samples studied; these disappear as the sample
size is increased. On the other hand, the critical temperature ($T_C$)
predicted by MFT are significantly higher than those obtained by MC
simulations (as would be expected).  Part of the difference in $T_C$
between the two methods is actually due to the fact that the Mn spins
in the MC simulations are taken to be classical variables, and quantum
operators in the MFA.  If we use classical Mn spins in MFA, we find
$T_C$ reduced by a factor of $\sim 2$. The remaining reduction is
presumably due to the usual neglect of fluctuations in MFA, which is
properly captured in MC simulations.

The striking feature, common to both results, is that the
magnetization curves have unusual shapes - linear or concave
upward. This is qualitatively similar to those found for the II-VI DMS
(Fig. \ref{fig4}), and what has been seen in experiments,\cite{Esch,Besch}
but very different from the convex upward $M(T)$ of conventional
ferromagnets (Fig.  \ref{fig2}). Again, as in the II-VI, the
magnetization does not reach its saturation value until very low
temperatures. Concurrently, the specific heat curves also exhibit a
peak at temperatures much lower than the critical temperature,
reflecting the entropy of the disordered spins present in the system
down to these low temperatures.\cite{MB}

By looking at the magnetization profile around the $T_C$, long-range
ferromagnetism in the disordered sample of III-V DMS is seen to appear
when a percolated cluster of polarized Mn spins is formed. However,
unlike in the spin-only model based on isolated hydrogenic centers
used for II-VI DMS in the previous section, the holes are {\em
delocalized} within this cluster for the parameters appropriate for
the III-V based DMS. Since the holes can more effectively minimize
their kinetic energy when maintaining the direction of their spin
during hopping, this delocalization of the holes within the percolated
cluster provides a very effective mechanism for alignment of all Mn
spins within the cluster in the same direction. This kinetic-induced
alignment mechanism is much more effective than mechanisms of
alignment of nearby BMPs  in insulating II-VI DMS, suggesting higher
critical temperatures in this case. Other reasons for enhancement of
critical temperatures in III-V DMS include the peaking of the impurity
wavefunctions at the Mn sites in this case where Mn is also the
dopant, and the ability of carriers in the compensated case to choose
states with wave-functions peaked in the regions with
higher-than-average Mn concentrations - the higher probability of
finding the holes in these regions leads to enhanced effective
interactions with the Mn spins. When all these factors are included,
we find indeed that the striking differences in critical temperatures,
by two orders of magnitude, in the two systems can be comfortably
explained, at least within MFA.

\subsection{Effect of disorder}

\begin{figure}
\centering
\parbox{0.45\textwidth}{\psfig{figure=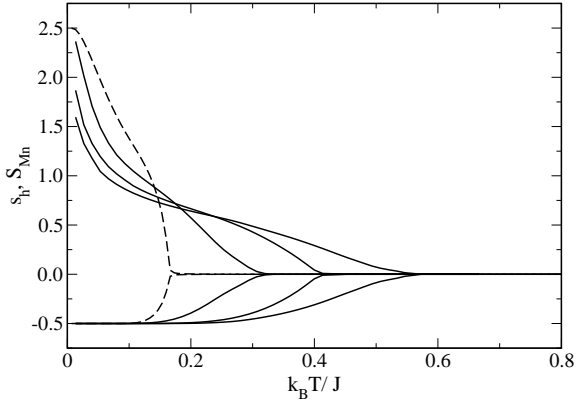,width=85mm,angle=270}}
\vspace{5mm}
\caption{\label{figMB4} The average Mn spin $S_{Mn}$ and average spin
per hole $s_{h}$ for doping concentration $x=0.00926$ and $p=10\%$. In
increasing order of $T_C$, the curves correspond to ordered, weakly
disordered, moderately disordered and completely random distributions
of Mn (see text). }
\end{figure}

Within the mean-field approximation, positional disorder in the Mn
spins for III-V DMS leads to a significant increase of the critical
temperature. Mn are the charged dopants in this case, so the situation
again appears to be similar to that in II-VI DMS.  Typical
magnetization curves obtained using MFA are shown in
Fig. \ref{figMB4}, for a doping $x=0.0093$ and $p=10\%$. In order of
increasing $T_C$, the four curves correspond to increasing disorder in
the positions of the Mn impurities. We start with a fully ordered,
simple cubic superlattice of Mn impurities inside the host
semiconductor (for this concentration, the superlattice lattice
constant is equal to three lattice constants of the underlying Ga FCC
sublattice). The corresponding average spins of the Mn and charge
carriers are shown by dashed lines in Fig. \ref{figMB4}. Then, we
introduce positional disorder of the Mn ions on the underlying Ga FCC
sublattice in varying amounts - (i) low-disorder where Mn spins are
randomly place on any of the nearest-neighbor sites of the original
superlattice sites; (ii) moderate disorder - where the Mn spins are
allowed to occupy any sites on the Ga sublattice, as long as the
distance between any two Mn is larger than two lattice constants; and
(iii) completely random positions of the Mn spins on the Ga FCC
sublattice of the host semiconductor.

In a fully ordered III-V DMS, below $T_C$ each Mn spin is equally
polarized, since translational invariance implies that the holes are
equally distributed among the various Mn sites and therefore create
the same effective magnetic field for each Mn spin.  [ In this respect
the ordered lattice for III-V is different from the situation
encountered for the ordered superlattice of charged dopants in the
II-VI DMS. In the II-VI, below $T_C$ the Mn spins inside the BMPs are
strongly polarized, while the Mn spins outside the BMPs are
practically unpolarized ]. The reason why $T_C$ is larger in a
disordered III-V DMS than an ordered one, is that the hole
wave-functions are pulled-in the regions with higher-than-average Mn
concentrations, where they can more effectively minimize their total
energy. The increased probability of finding the holes in this smaller
volume occupied by the cluster leads to effectively larger couplings
$J_{eff}$ of the Mn spins in the cluster,\cite{MB} and therefore
increased critical temperatures. In other words, in the disordered
III-V DMS the holes only need to polarize a smaller fraction of the Mn
spins in the system and get polarized in their own term. In an ordered
Mn sample, the holes polarize equally {\em all} the Mn spins in the
system, and this can only happen at rather low temperatures, given the
small number of holes as compared with the number of Mn spins. While
MFA shows a strong dependence of $T_C$ on disorder, this is likely to
be modified once fluctuation effects left out in MFA are included, as
in a MC simulation.

\begin{figure}[b]
\centering
\parbox{0.45\textwidth}{\psfig{figure=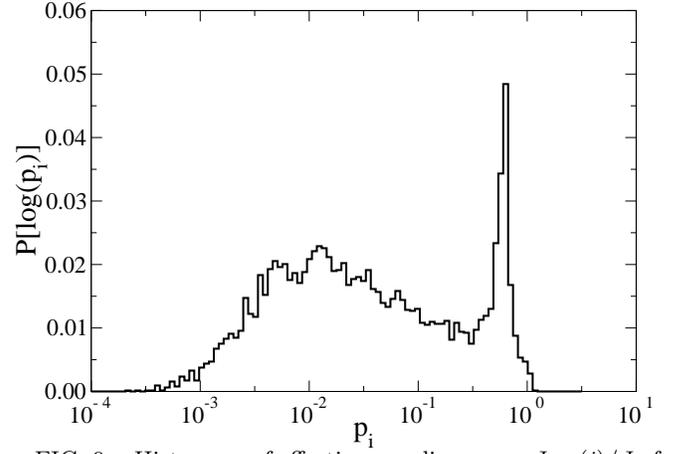,width=85mm,angle=270}}
\caption{\label{figMC2} Histogram of effective couplings $p_i=J_{eff}(i)/J$
of different Mn spins at $k_B T /J =0.01$, for x=0.01 and relative
hole to Mn concentration $p=10\%$. This distribution was found using
Monte Carlo simulations.\cite{malcolm} }
\end{figure}

The unusual shape of the magnetization curves is a consequence of the
relatively small number of charge carriers as compared to the number
of Mn spins. In a disordered system, we can identify two types of Mn
spins: strongly-interacting Mn spins from the percolated cluster,
which polarize at high temperatures and lead to the ferromagnetic
transition at $T_C$, and weakly-interacting Mn spins from the regions
outside the percolated cluster. Since these outside regions have low
hole density in our model, the effective coupling of their Mn spins
(which is proportional to the probability of finding holes nearby) is
rather small, Consequently, these spins do not polarize unless the
temperature is comparable in size to their effective coupling. We have
used this picture to obtain a simplified but fairly accurate
description of the magnetic and thermodynamic properties of the DMS
based on a two-component model.\cite{malcolmTF} We start from a
histogram of the effective couplings $J_{eff}$ of all the Mn spins,
obtained by averaging over many realizations of disorder. Such a
histogram of $J_{eff}/J$ obtained using Monte Carlo simulations for
$x=0.01$ and $p=10\%$ is shown in Fig. \ref{figMC2}. As can be seen,
it is a very wide distribution, from very large $J_{eff} \sim J$ for
strongly interacting Mn spins, to extremely small $J_{eff}/J \sim
10^{-3}$ values for weakly interacting Mn spins. Histograms obtained
within the mean-field approximation have very similar shapes, except
that their width is even larger.\cite{MB}

For such wide distributions, at any given temperature $k_BT$, we
divide the spins into weakly/strongly-interacting categories,
depending of whether their effective coupling $J_{eff}$ is
smaller/larger than $\gamma k_B T$. Then, we replace the complex
distribution of couplings shown in Fig. \ref{figMC2} by two
$\delta$-functions representing the two spin components. The values of
the nominal couplings $J_1$ and $J_2$ of the
weakly/strongly-interacting spin components are simply the average of
all the couplings of weakly/strongly interacting spins.  The constant
$\gamma$ is found from a fit of, for instance, the magnetization curve
provided by this simplified model. Other thermodynamic quantities,
such as susceptibility and specific heat are then shown to be quite
well described by this simple model.\cite{malcolmTF} In contrast, we
have verified that replacing all the couplings by a {\em single}
coupling corresponding to the average over the entire distribution
leads to curves very different than the ones obtained with the
original distribution.  We believe this simplified model could provide
a simple tool for interpretation of experimental curves. So far, most
attempts have been to try to fit the magnetization curves (for
instance) with only one coupling. While this may recapture part of the
curve near and below $T_C$, \cite{Ohnorev} it turns out that it only
accounts for a rather small percentage of the total number of Mn spins
expected to be in the system. This suggests that a second component is
missing. In fact, fits in terms of two components, one ferromagnetic
and one paramagnetic, have already been performed in order to explain
the shapes of the measured $M(H,T)$ curves.\cite{oiwa}

\begin{figure}
\centering
\parbox{0.45\textwidth}{\psfig{figure=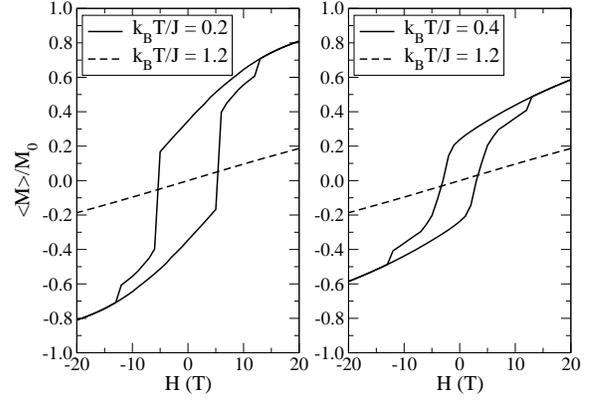,width=85mm,angle=270}}
\caption{\label{his} Hysteresis curves obtained within the mean-field
approximation for {\em one disorder realization} corresponding to a Mn
concentration $x=0.03$ and hole to Mn ratio $p=10\%$ (corresponding to
a critical temperature $k_BT_C/J=0.85$). Averages over several disorder
realizations are needed to obtain smooth curves.  }
\end{figure}

In a conventional ferromagnet, an external magnetic field will lead to
a fast increase of the magnetization from its value in the absence of
the field, to the saturation value $M_0=N g \mu_B S$, where $N$ is the
concentration of spins $S$ in the system. A hysteresis curve
associated with the existence of ferromagnetism below $T_C$ is also
observed. In III-V DMS samples, the hysteresis curves are clearly
observed as well. However, even at rather large fields $H$, $M(H)$
does not saturate, but continues to increase with increasing magnetic
field. This feature has been attributed to a ``paramagnetic''
component,\cite{oiwa} and it obviously corresponds to the
weakly-interacting component of nearly-free spins of our simplified
two-component model. In fact, we have generated $M(H,T)$ curves within
the mean-field approximation, and these features are clearly present
(see Fig. \ref{his}), in qualitative agreement with measurements.

\subsection{Effect of other interactions}

We have investigated in detail the effect of the on-site disorder term
$u(i)$ and of the on-site Coulomb repulsion $U$ elsewhere.\cite{MB}
The on-site disorder is due to the Coulomb potential created by the
charge impurities responsible for compensation (such as As$^{++}$
antisites). We have considered two extreme possibilities. In the first
case, we assume that these potentials are completely un-correlated,
and model them by choosing random values for $u(i)$ within an interval
$[-W, W]$. The estimate $W\approx$ 1 Ry is obtained following standard
considerations for doped semiconductors.\cite{SE} In the second case,
we attempt a simple modeling of the effect of As antisites. We choose
random positions for these As defects on the Ga sublattice and
identify their two nearest neighbor Mn sites.  Each such As impurity
has an effective charge $+2e$, and therefore will contribute an
on-site Coulomb potential $+2e^2/\epsilon r$ at a Mn impurity site
which is at a distance $r$ from it. However, since the Mn ions also
have effective ionic charge $-e$, the As potential is screened
(partially compensated) by the potential of the Mn impurities nearby
it. Therefore, we assume that each As antisite only contributes to the
on-site potential $u(i)$ of its two nearest Mn neighbors, with the
contribution to the other Mn sites being screened out by the
contribution of these two nearest Mn sites.  The presence of the
charged impurities responsible for compensation increases the amount
of disorder (inhomogeneity) in the system, since the holes will avoid
the regions were these defects are located. Thus, one might assume
that $u(i) \ne 0$ will lead to a further increase of $T_C$. However,
in fact we find a decrease of $T_C$ for these models of compensation,
especially for the second model.\cite{MB} This is a consequence of the
fact that due to the presence of nearby As antisites, holes now avoid
some Mn sites that would otherwise be part of dense clusters. Thus,
the system effectively moves towards the more homogeneous regime, with
lower $T_C$.  An opposite effect is provided by the on-site
electron-electron Coulomb repulsion, the presence of which was found
to lead to an increase of $T_C$, since it aids in the splitting of the
up and down spin bands, favoring spin polarization at higher
temperatures.\cite{MB}

A quantitative determination of the effects of these types of
interactions will have to wait until more details are known about the
compensation processes. A theory that properly and self-consistently
describes the screening processes is also necessary.

\section{Concluding Remarks}
\label{sec5}

In this paper, we have discussed the behavior of a model of DMS in the
low density regime, based on a simple tight-binding hydrogenic model
of the impurity band. Such a model takes into account, at the very
outset, the inherent disorder present in the experimental system,
namely the random position of the dopants. Other
models\cite{MacDonald-Dietl} start from an electron gas exhibiting the
translational symmetry of the host lattice and ignore the disorder of
the alloy system.  While the latter may be the appropriate starting point
for the high carrier density regime, it does not allow for a
metal-insulator transition, and consequently misses the unusual
transport and magnetic behavior associated even with metallic systems
in the vicinity of such a transition.  In contrast, our model starts
from the low density insulating side, and at least for conventional
doped semiconductors, has been found to be applicable to densities up
to a factor of ~3 above the metal-insulator transition.\cite{Mott}
 
For the case of II-VI DMS, we have restricted ourselves to low
densities corresponding to the insulating phase, for the case of a
half-filled band {\em i.e.,} no compensation. In this limit, a
spin-only description of the bound carriers is appropriate.  [We note,
however, that such a spin only description has been very successful
for the low temperature thermodynamic and magnetic properties in
conventional doped semiconductors, both uncompensated and compensated,
for densities up to the metal-insulator transition\cite{Bhatt} and
even somewhat into the metallic phase\cite{Bhatt2,Hirsch} provided an
itinerant Fermi-liquid like second component is added to the
description of these highly disordered systems].  For III-V DMS, where
large compensation is found to be experimentally present, presumably
due to antisite defects, we have adopted a full fermionic description
of the carriers. Such a model allows for both an insulating and a
metallic phase. However, as explained in the body of the paper, the
model we have studied is simplified, and neglects several terms in the
full many-body Hamiltonian describing these complicated materials.

Despite the rather different model descriptions (spin vs.  fermion)
for the two cases, as well as methods of solution (Monte Carlo
vs. Mean Field Approach), we find a remarkable similarity in the
qualitative predictions concerning the magnetic and thermodynamic
properties.  Most striking are the unusual magnetization curves
$M(T)$, with linear to concave upwards shape over much of the
ferromagnetic region, in striking contrast to conventional uniform
ferromagnets.  This appears to be a combined result of low carrier
density and strong disorder. As a consequence, the ferromagnetic
transition has percolation like characteristics, with only a small
fraction of the material carrying the bulk of the ferromagnetism
around $T_C$. The remaining portion of the material orders gradually
as the temperature is lowered, and unlike in most conventional
ferromagnets, saturation magnetization is not reached until well below
$T_C$.  Such an inhomogeneous magnetization results in unusual
susceptibility and specific heat in the low temperature ordered phase,
and would imply substantial inhomogeneities in the local field at Mn
sites, which could be probed, {\em e.g.,} by NMR measurements. Unusual
hysteresis curves in $M(H)$ below $T_C$ are also implied, with
saturation occurring well beyond where the loops close.  We have
checked for the case of the III-V DMS that within a simple impurity
band description, these effects are robust.\cite{MB} However, as the
carrier density is increased (by reducing the compensation, or raising
the Mn concentration), the anomalous shape of $M(T)$ becomes less
prominent: $M(T)$ assumes the convex upward shape of conventional
uniform ferromagnets, and the ensuing unusual properties discussed
above gradually fade away.

In contrast to the qualitative shape of the magnetization curves and
the ensuing thermodynamic and magnetic behavior, the actual transition
temperatures of the two systems are known to be rather different (
from a few degrees kelvin\cite{bookIIVI} for the II-VI DMS, to several
hundreds\cite{Ohnorev,Esch,Hebard,Sono} for the III-V DMS ). Certainly
one reason for this difference is the increased weight of the hole
wave-function at the cation (II/III) site where the Mn spin resides in
the III-V semiconductors relative to the II-VI semiconductors, as may
be seen from a tight-binding description of valence bands\cite{Chadi}
of zinc-blende structure semiconductors.  However, an additional
reason within an impurity band description of carriers, is that in
III-V DMS the Mn sites are centers of the impurity wave-functions,
while for II-VI DMS, the carrier impurity sites are distinct from the
Mn. Consequently, the Mn sites see a lower amplitude of the carrier
wave-function, and thus a lower effective exchange coupling in the
II-VI.  This peaking of the impurity wave-function at the Mn site in
the III-V based DMS, leads to a further enhancement of their $T_C$
vis-a-vis the II-VI based DMS.

For both the II-VI and the III-V DMS, we find that $T_C$ is enhanced
by disorder. This can be understood by recognizing that in a heavily
disordered system, nature is able to create global ordering by finding
the tortuous percolative pathway necessary when the average
coordination number is much below that of any uniform
lattice.\cite{SE} In III-V, the large compensation adds an additional
degree of freedom to the carriers, in the choice of amplitudes on
different sites, which again leads to wide variation in the effective
fields at different sites, and implies a percolative aspect to the
magnetic ordering transition. In mean field, we find the enhancement
of $T_C$ to be quite large; however, preliminary Monte Carlo results
suggest lower effects of disorder on $T_C$ than given by the
mean-field approach.\cite{malcolm}

Disorder effects on the electronic wave-functions will lead to
significant transport anomalies, especially near the metal insulator
transition, as has been seen experimentally.\cite{Ohnorev} It will
also likely affect the nature and amount of magnetic scattering of
carriers injected into the system. While the naive expectation is that
disorder should increase spin-flip scattering, it may be significantly
reduced for carriers near the Fermi level. This is because we find
that these states have large amplitudes along the percolating backbone
of the system, where the Mn moments are magnetized well above the
average global magnetization.  In this regime, many standard models
devised for translationally invariant systems ({\em e.g.,}
relationship of anomalous Hall effect to bulk magnetization) may not
be applicable, and such interpretations should be used with care.

One other approximation inherent in our work is the assumption that
electron and hole doping give rise to Hamiltonians that are
qualitatively similar, though quantitatively different (holes have
angular momentum 3/2, while electrons have spin 1/2). This is what is
found for free holes:\cite{MacDonald} although the more complicated
anisotropic wave-functions for holes leads to quantitative
differences, qualitatively the results are similar, in that both
systems lead to ferromagnetic ordering. Recently, however, since the
Spintronics 2001 conference, it has been proposed\cite{Janko} that
spin-orbit coupling can lead to effective spin-spin couplings that are
{\em anisotropic} in spin space, and the positional disorder
effectively leads to random anisotropy. This could lead to frustration
effects not present in our model, and if true, would need to be put in
for hole doped systems to achieve full understanding of magnetic
ordering and carrier transport in DMS systems.

Finally, we discuss the applicability of our results to actual III-V
DMS in the regime of large $T_C$. While our model is based on the
insulating, low density limit, how many of its features persist into
the metallic phase at higher densities and temperatures, is dependent
on the nature of the filled electronic states at temperatures $T_C$
and below. In the model we have studied, the host valence bend is
completely neglected, and its inclusion is not expected to lead to
qualitative changes, because it lies several hundred meV above the
Fermi level. However, this is a consequence of an impurity band with a
density of states that is characteristic of a bandwidth of order
hundred meV also. Should the impurity band become much broader in the
actual system due to effects we have left out, it will likely merge
into the host valence band, and the states will be strongly
mixed. Nevertheless, the occupied states for small filling (low Mn
density and large compensation) would have significant effects of
disorder. This, in turn, implies that the anomalous behavior exhibited
by our model would be present, but with lower magnitude than shown by
our calculations. The clearest signature of these would likely come
from local probes, which would be able to determine the distribution
of {\em local} fields, and hence {\em local} density of states at
various sites. Such input into phenomenological models should provide
a fruitful avenue for a more in-depth study of the fascinating world
of real Diluted Magnetic Semiconductors, which offer both a
significant promise in terms of their applications in spintronics, and
a challenge in terms of their fundamental understanding.

\section*{Acknowledgments}
This research was supported by NSF DMR-9809483.  R.N.B. acknowledges
the hospitality of the Aspen Center of Physics during their Spins in
Nanostructures workshop, when this presentation for the Spintronics
2001 conference was organized.  M.B. was supported in part by a
Postdoctoral Fellowship from the Natural Sciences and Engineering
Research Council of Canada.

\end{document}